\documentclass[aip,reprint]{revtex4-1}
\usepackage{amsmath,amsthm,amsfonts,amssymb,amscd,bbold,mathrsfs,bm}
\usepackage{graphicx}
\usepackage{dcolumn}
\usepackage{physics}
\usepackage{hyperref}
\usepackage{upgreek}

\begin{document}
\title{Cancelling second order frequency shifts in Ge hole spin qubits via bichromatic control}
\author{Xiangjun Tan}
\affiliation{Department of Physics and Astronomy, University College London, WC1E 6BT London, United Kingdom}
\author{Zhanning Wang}
\affiliation{Instituto de Ciencia de Materiales de Madrid, Consejo Superior de Investigaciones Científicas, Sor Juana Inés de la Cruz 3, 28049 Madrid, Spain}
\author{Wenkai Bai}
\affiliation{Shaanxi Key Laboratory for Theoretical Physics Frontiers, Institute of Modern Physics, Northwest University, Xi'an 710127, China}
\affiliation{Institute of Physics, Chinese Academy of Sciences, Beijing 100190, China}
\author{Hanjie Zhu}
\affiliation{Center for Joint Quantum Studies and Department of Physics, School of Science, Tianjin University, Tianjin 300350, China}
\affiliation{Institute of Physics, Chinese Academy of Sciences, Beijing 100190, China}
\date{\today}

\begin{abstract}
Germanium quantum dot hole spin qubits are compatible with fully electrical control and are progressing toward multi-qubit operations.
However, their coherence and calibration stability are limited by charge noise and drive-induced frequency shifts.
Here we theoretically demonstrate that a bichromatic driving scheme cancels the second order frequency shift from the control field without sacrificing the electric dipole spin resonance (EDSR) rate, and without additional gate design or microwave engineering.
Based on this property, we further demonstrate that bichromatic control creates a wide operating window that compensates static charge-induced resonance offsets during fixed-frequency operation.
This method provides a low-power route to a stabler frequency operation in germanium hole spin qubits and is readily transferable to other semiconductor spin qubit platforms.
\end{abstract}
\maketitle

The family of semiconductor quantum dot hole spin-orbit qubits in the germanium platform has drawn much attention in recent years for its relatively long coherence time and potential for industrial level production \cite{Watzinger2018,Scappucci2021,Fang2022,Burkard2023,James2025}, in line with rapidly growing impurity-free fabrication \cite{Sammak2019,Stehouwer2023,Corley-Wiciak2023}, precise external field control \cite{Ke2022,Camenzind2022,Floor2024}, and high fidelity readout and error correction protocols \cite{Hendrickx2021,Froning2021_NN,Geyer2024,Yujun2025}.
The inherent strong spin-orbit coupling (SOC) in germanium hole systems, due to the valence band light-hole-heavy-hole (LH-HH) coupling \cite{Rashba1988,Winkler2002,Kloeffel2011,Durnev2014,Marcellina2018,Bosco2021,Martinez2022,Carlos2022,Bosco2022,Adelsberger20222,Vecchio2023,Lidal2023,Secchi2024,Andrea2025,Rotaru2025,Vecchio2025,Kelvin2025}, allows for purely electrical manipulation of the spin states and EDSR \cite{Golovach2006,Bulaev2007,Ares2013,Milivojevic2021,Fanucchi2025}.
Due to the natural p-type band structure, the contact hyperfine interaction is suppressed, and with the support of isotopic purification in group IV systems, recent breakthroughs have extended the dephasing time $T_2^*$ to $20~\upmu$s in Ge with a $B^{-1}$ scaling law \cite{Stehouwer2025}.
Furthermore, owing to sensitive $g$-tensor modulations and anisotropic orbital magnetic field terms, studies have shown that an optimal dephasing-time region can be achieved by tuning the external magnetic field, the gate electric field, and applied strain \cite{Crippa2018,Niquet2018,Wang2021,Jirovec2022,Abadillo2023,Michal2023,Mena2023,Shalak2023}.
Recent advances have also highlighted the possibility of integrating hole spin qubits with circuit QED systems, which favour dispersive readout of the qubit states and provide a bridge to multi-qubit entanglement \cite{Kloeffel2013,Bosco2022_PRL,Xiangjun2025,tan2025}.

On the other hand, the strong SOC also couples the qubit to charge defects, and ensemble $1/f$ type electric noise, together with hyperfine interactions from isotopes, limit the coherence of the qubit in Ge hole nanostructures \cite{Wang2021,Mauro2024,Shalak2023,Sina2025}.
As the Ge quantum dot array scales, suppressing frequency drift at the individual qubit level becomes increasingly important for extending the dephasing time and gate fidelities \cite{Hendrickx2021,Hendrickx2020_NC,Hendrickx2024}.

Recent work has explored bichromatic driving to improve addressability and mitigate spectral crowding in quantum dot arrays, enabling multi-qubit control \cite{Gyorgy2022,John2024}.
We identify a complementary single qubit benefit: with the primary tone resonant for a fast one photon EDSR gate, a detuned auxiliary tone acts as an independent dressing knob that cancels the net second order drive-induced frequency shift.
This auxiliary tone also compensates static charge-induced detuning offsets, thus improves spin qubit control fidelity, mitigates recalibration overhead, where high-fidelity operation often requires frequency feedback to keep the control tone on resonance \cite{Takeda2018,Capannelli2025,Tanttu2024,Hendrickx2024}.

We first identify the HH qubit subspace by numerically diagonalizing the full quantum dot Hamiltonian, and then use a Floquet-Magnus expansion to derive the bichromatic drive-induced second order shifts.
We find (i) a broad parameter window where the EDSR-induced frequency shift can be cancelled with low power dissipation, and (ii) a broad regime in which the residual detuning can be minimized relative to monochromatic driving at the same gate speed; for example, at $B_x=1$~T, $E_{\text{gate}}=10$~MV/m, and $E_{\text{AC}}=10$~kV/m, we find an approximately threefold reduction in the minimum residual detuning magnitude $|\delta\omega_{\mathrm{res}}|$.
Although we focus on germanium, the approach should generalize to p-type silicon MOS system \cite{Jehl2016,Liles2024}.

\begin{table}[htbp!]
\begin{ruledtabular}
\begin{tabular}{lcc}
Label & value & unit \\\hline
$\gamma_1, \gamma_2, \gamma_3$ & $13.14, 4.59, 5.13$ & - \\
$a_v, b_v$ & $2.0, -2.3$ & eV \\
$n$ & 0.25 & - \\
$\epsilon_{xx}, \epsilon_{yy}, \epsilon_{zz}$ & $-0.6\%, -0.6\%, 0.42\%$ & - \\
$\kappa, q$ & $3.14, 0.07$ & -\\
$a_x, a_y, L$ & $50, 50, 15$ & nm \\
$\bar{\gamma}$, $\delta$ & $(\gamma_3 +\gamma_2)/2$, $(\gamma_3 -\gamma_2)/2$ & -
\end{tabular}
\end{ruledtabular}
\caption{
List of parameters and their values.
Here, $\gamma_1, \gamma_2, \gamma_3$ are Luttinger parameters for Ge.
$a_v, b_v$ are the deformation constants.
$n$ is the composition fraction of Si in Si$_{n}$Ge$_{1-n}$ layer, which determinants the $\epsilon_{xx} = \epsilon_{yy}$ and $\epsilon_{zz}$ in the biaxial strain through the Poisson ratio \cite{Pollak1968, Terrazos2021}.
$a_x, a_y$ and $L$ are the in-plane quantum dot size and quantum well width respectively \cite{Hendrickx2018,Hendrickx2021,Hendrickx2024}.}
\label{Table: params}
\end{table}

We consider a single germanium HH qubit in a SiGe/Ge/SiGe heterostructure, where the static biaxial strain is set by the composition fraction in Table~\ref{Table: params}.
The total Hamiltonian is $H_{\text{total}} = H_{\text{LK}} + H_{\text{BP}} + H_{\text{Conf}} + H_{\text{Zeeman}}$.
We use a four band Luttinger-Kohn Hamiltonian and neglect the large split-off band gap due to the large split-off gap ($\Delta_{\text{SO}}=290$~meV) \cite{Luttinger1955, supp}:
\begin{equation}
\begin{aligned}
H_{\text{LK}}= & \frac{\hbar^2}{2 m_0}\left[\left(\gamma_1+5\gamma_2/2\right) \bm{k}^2-2 \gamma_2 \bm{k}^2\cdot\bm{J}^2\right. \\
& \left.-4 \gamma_3\left(\{k_x, k_y\}\{J_x, J_y\}+\text {c.p.}\right)\right] 
\end{aligned}
\end{equation}
We apply the Peierls substitution $\bm{k} \to \bm{k} + e \bm{A}/\hbar$ in the gauge $\bm{A} = -\bm{x} \times \bm{B}/2$, where $\bm{B}$ is the static magnetic field \cite{Bosco2021}.
In this work, we only consider the static biaxial strain described by the Bir-Pikus Hamiltonian, $H_{\text{BP}} = \operatorname{diag}[P_\epsilon + Q_\epsilon, P_\epsilon + Q_\epsilon, P_\epsilon - Q_\epsilon, P_\epsilon - Q_\epsilon]$, where $P_\epsilon = -a_v (\epsilon_{xx} + \epsilon_{yy} + \epsilon_{zz})$, $Q_\epsilon = -b_v (\epsilon_{xx} + \epsilon_{yy} - 2 \epsilon_{zz})/2$.
All parameters can be found in Table~\ref{Table: params}.
While full device-level modelling is important \cite{Shalak2023, Abhikbrata2025}, we use this minimal model to focus on cancelling the second order frequency shift.
We neglect inhomogeneous or shear strain, which can enhance the EDSR Rabi rate via $g$-tensor modulation and stronger SOCs \cite{Crippa2018,Abadillo2023,Mena2023}, and we likewise ignore the interface roughness and associated Dresselhaus-type SOC \cite{Culcer2010,Boross2016,Niquet2019,Corley-Wiciak2023,Carlos2022,Wang2024}.
The confinement potential $H_{\text{Conf}}$ consists of a perpendicular gate electric field along the z-axis and two parabolic potentials along the x- and y-axes:
\begin{equation}
\frac{m_{\text{HP}}}{2} \left( \omega_{0,x}^2 x^2 + \omega_{0,y}^2 y^2 \right) +
\begin{cases}
e E_{\text{gate}} z, & z \in \left[-\frac{L}{2}, \frac{L}{2}\right], \\
\infty, & \text{elsewhere.}
\end{cases}
\end{equation}
Here, $m_{\text{HP}} = m_0/(\gamma_1+\gamma_2)$ is the in-plane effective mass.
The confinement frequencies $\omega_{0,x}$ and $\omega_{0,y}$ set the effective quantum dot size (independent of $E_{\text{gate}}$ in this model), which can be further modified by the magnetic field \cite{Bopp2025}.
$E_{\text{gate}}$ is assumed uniform, and we neglect wavefunction leakage into the SiGe layer \cite{Terrazos2021, Jiawei2025}.
The Zeeman Hamiltonian is $H_{\text{Zeeman}} = 2 \mu_B \left(\kappa \bm{J}+q \bm{J}_3\right) \cdot \boldsymbol{B}$, where $\mu_B$ is the Bohr magneton, $\bm{J} = (J_x, J_y, J_z)$ is the total angular momentum operator, and $\bm{J}_3 = (J_x^3, J_y^3, J_z^3)$ accounts for the cubic anisotropic contribution.

To obtain the qubit subspace and eigenstates, we use the exact diagonalization method \cite{Abhik2023}.
The basis combines infinite square well states along the z-axis and harmonic oscillator state along the x- and y-axes, respectively: $\ket{\psi_{n_x}, \psi_{n_y}, \psi_{n_z}, \chi}$, where $\chi$ denotes the HH-LH spinor.
All the products of two canonical wave vector operators are symmetrized as $k_i k_j \to \{k_i, k_j\}/2$.
For each working point of $E_{\text{gate}}$ and $\bm{B}$, we diagonalize $H_{\text{total}}$ to define the qubit subspace, spanned by the ground and the first excited eigenstates, which forms a Kramers pair at $\bm{B}=0$.
We have verified that the next orbital excited state is adequately gapped across all the working points.
The ground state and the first excited states are labelled as $\ket{\mathbb{1}}$ and $\ket{\mathbb{2}}$ with energies $E_{\mathbb{1}}$ and $E_{\mathbb{2}}$, defining the qubit Larmor frequency $\omega_0=(E_{\mathbb{2}}-E_{\mathbb{1}})/\hbar$.
Higher excited states are denoted as $\ket{\mathbb{n}}$ with energy $E_{\mathbb{n}}$.

The bichromatic driving term is $H_{\text{AC}} = e E_1 x \cos(\omega_1 t + \phi_1) + e E_2 x \cos(\omega_2 t + \phi_2)$, where $E_{1,2}$ are the in-plane electric field amplitudes, $\omega_{1,2}$ the driving frequencies, and $\phi_{1,2}$ the phases.
The qubit-field coupling follows a dipole interaction model; SOC effects, HH-LH mixing, and orbital magnetic field effects are carried by the qubit states obtained by exact diagonalization of $H_{\text{total}}$.  
Now, the driven qubit Hamiltonian $H_{\text{qubit}}$ is:
\begin{equation}
H_{\text{qubit}} = \frac{\hbar \omega_0}{2} \sigma_z+
\sum_{\alpha} \frac{\hbar}{2} \Omega_{\alpha} ( \mathrm{e}^{\mathrm{i} \theta_\alpha} \sigma_+ + \mathrm{e}^{-\mathrm{i} \theta_\alpha} \sigma_-) \cos(\omega_\alpha t + \phi_\alpha) \,.
\end{equation}
The index $\alpha \in \{1,2\}$ indicates the driving frequencies, and $\Omega_{\alpha} = e E_{\alpha}|\mel{\mathbb{1}}{x}{\mathbb{2}}|/\hbar$; the phase angle is encoded in $\mathrm{e}^{\mathrm{i} \theta_\alpha}$.  
Here, $\sigma_+ = (\sigma_x + \mathrm{i} \sigma_y)/2$, $\sigma_- = (\sigma_x - \mathrm{i} \sigma_y)/2$, where $\sigma_{x,y,z}$ are Pauli matrices.  
Beyond enabling coherent manipulation of the spin states, these driving terms also produce a second order frequency shift $\delta \omega^{(2)}$ arising from two channels: the Bloch-Siegert shift (counter-rotating terms) and the AC Stark shift (near-resonant virtual transitions via states of $H_{\text{total}}$).
We evaluate both within a Floquet-Magnus perturbative treatment as second order corrections to the quasi-energy.
Following the discussion above, we neglect driving-induced sideband modulations, i.e., the beat-note envelope and Floquet satellite transitions produced by the two-tone drive (satellite peaks around the central resonance) \cite{Shirley1965,Grifoni1998}.
Since we focus on the second order quasi-energy shift of the central transition at $\omega_1 \simeq \omega_0$, these sideband terms only redistribute spectral weight into satellites and, to the perturbative order considered (and under long time or lock-in averaging), do not shift the central resonance frequency.

We first evaluate the AC Stark shift from higher excited states.
In general, we construct a unitary transformation $U(t) = \mathrm{e}^{S(t)}$, where $S(t)$ is an anti-Hermitian operator, used to transform $H_{\text{qubit}}$ to second order: $H_{\text{eff}} = U(t) H_{\text{qubit}} U^\dagger(t) + \mathrm{i}\hbar \dot{U}(t) U^\dagger(t)$.
Since there are only two frequency components, $H_{\text{AC}}$ can be rewritten as:
\begin{equation}
H_{\text{AC}} = \frac{ex}{2}\sum_{\alpha} E_\alpha (\mathrm{e}^{\mathrm{i}\phi_\alpha} \mathrm{e}^{\mathrm{i}\omega_\alpha t} +
\mathrm{e}^{-\mathrm{i}\phi_\alpha} \mathrm{e}^{-\mathrm{i}\omega_\alpha t}) \,.
\end{equation}
Expanding the full Hamiltonian and taking the time average, the diagonal correction to second order is given by:
\begin{equation}
E_{\text{AC},\mathbb{m}}^{(2)} = \sum_{\mathbb{n}>2} \sum_{\alpha} \frac{e^2 E_\alpha^2}{4} |\mel{\mathbb{m}}{x}{\mathbb{n}}|^2 \frac{2\Delta_{\mathbb{m}\mathbb{n}}}{\Delta_{\mathbb{m}\mathbb{n}}^2 - \hbar^2 \omega_\alpha^2} \,,
\end{equation}
where the energy detuning $\Delta_{\mathbb{m}\mathbb{n}} = E_{\mathbb{m}} - E_{\mathbb{n}}$.
The index $n$ runs over all higher excited states from $\mathbb{n}=3$ to $\operatorname{dim}(H_{\text{total}})$, thereby accounting for all transition channels.

Focusing on the qubit subspace, using a similar approach, the diagonal corrections from the counter-rotating and near-resonant parts ($\pm$ for $\ket{\mathbb{1}}$ and $\ket{\mathbb{2}}$, respectively) are:
\begin{equation}
E_{\text{BS}} = \sum_{\alpha} \frac{\pm \hbar \Omega^2_\alpha}{4(\omega_0+\omega_\alpha)}  \quad E_{\text{AC}} = \sum_{\alpha} \frac{\pm \hbar \Omega^2_\alpha}{4(\omega_0-\omega_\alpha)} \,.
\end{equation}
To check the corrections to the Rabi rate, the off-diagonal element is dynamically renormalized as:
\begin{equation}
\Omega_{\text{eff},\alpha} = \Omega_{\alpha} \left( 1 - \frac{\Omega^2_{\alpha}}{4(\omega_{0}+\omega)^2} \right) \,,
\end{equation}
which is negligible since $\Omega_{\alpha}$ (MHz) is much smaller then $\omega_0$ (GHz).
Hence we use the bare $\Omega_{\alpha}$ below.
Collecting the AC Stark and Bloch-Siegert contributions, the second order frequency shift can be written as $\delta \omega^{(2)} = C E_\alpha^2$, where the second order electric field response $C$ reads:
\begin{equation}
\begin{aligned}\label{Eq: response function}
{C} = & \frac{e^2}{4\hbar} \frac{|x_{\mathbb{1}\mathbb{2}}|^2}{\Delta_{\mathbb{2}\mathbb{1}} + \hbar \omega} 
+ \frac{e^2}{4\hbar} \frac{|x_{\mathbb{1}\mathbb{2}}|^2}{\Delta_{\mathbb{2}\mathbb{1}} - \hbar \omega} \\
& + \frac{e^2}{4\hbar} \sum_{\mathbb{n} >2} \left[ |x_{\mathbb{2}\mathbb{n}}|^2 f(\Delta_{\mathbb{2}\mathbb{n}},\omega)-|x_{\mathbb{1}\mathbb{n}}|^2 f(\Delta_{\mathbb{1}\mathbb{n}},\omega) \right] \,,
\end{aligned}
\end{equation}
where $|x_{\mathbb{m}\mathbb{n}}| = \mel{\mathbb{m}}{x}{\mathbb{n}}$, $\Delta_{\mathbb{m}\mathbb{n}} = E_{\mathbb{m}} - E_{\mathbb{n}}$, and $f(\Delta_{\mathbb{2}\mathbb{n}},\omega) = 2\Delta_{\mathbb{m}\mathbb{n}}/(\Delta_{\mathbb{m}\mathbb{n}}^2 - \hbar^2 \omega_\alpha^2)$.
We refer to the three terms in Eq.~\eqref{Eq: response function} as $C_{\text{AC,2}}$, $C_{\text{BS}}$, and $C_{\text{AC},1}$ respectively.

\begin{figure}[htbp!]
\centering
\includegraphics[width=\linewidth]{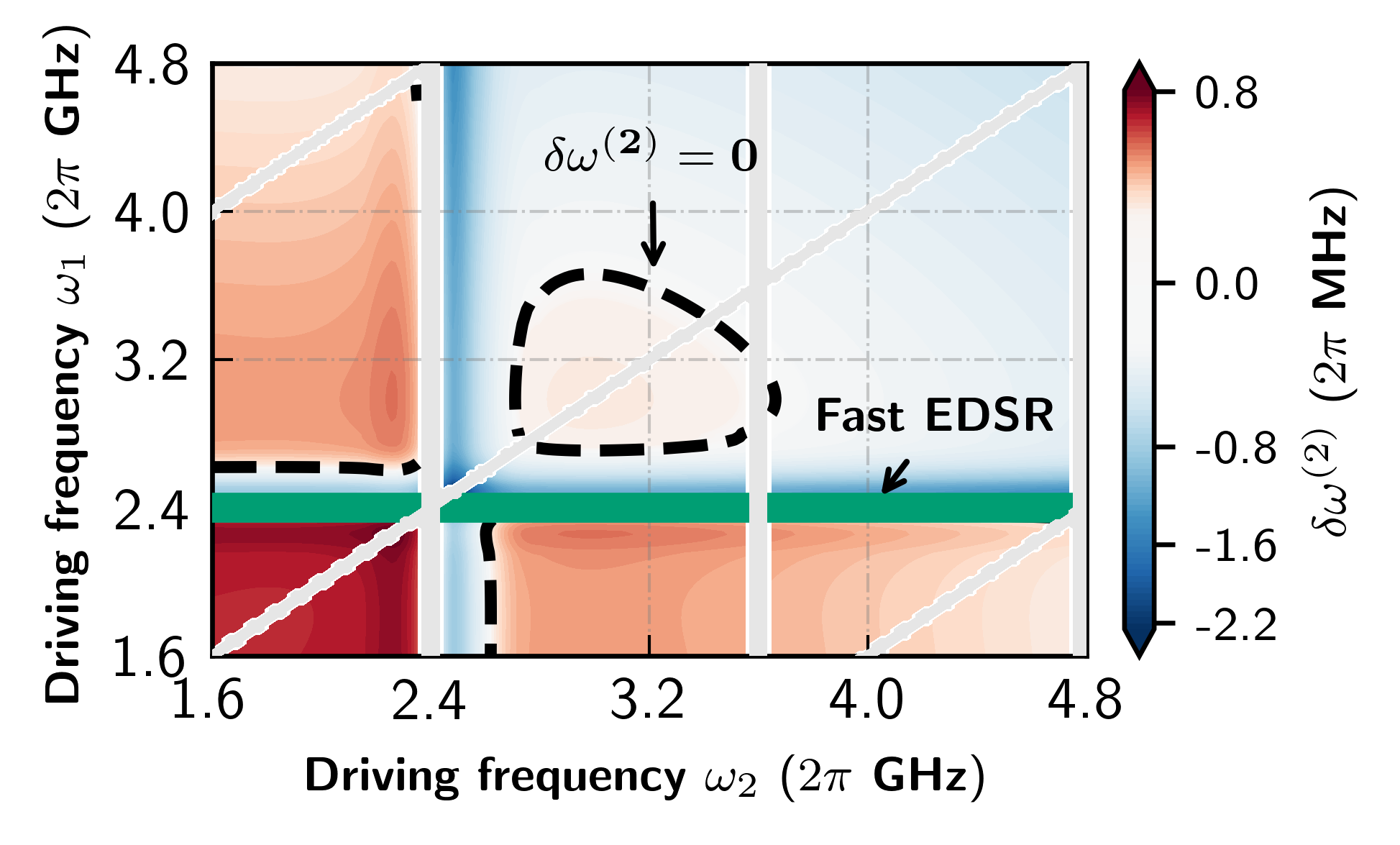}
\caption{
Heatmap of the second order frequency shift $\delta\omega^{(2)}$ as a function of the two driving frequencies $\omega_1$ and $\omega_2$.
The parameters used to generate this figure are $E_{\text{gate}}=10$~MV/m, $B_x=1$~T, and $E_1=E_2=10$~kV/m.
The dashed black contour indicates $\delta\omega^{(2)}=0$.
Grey shaded regions denote parameter regimes associated with multiphoton processes and are therefore excluded from the analysis.
The green highlighted region indicates the fast EDSR regime.}
\label{Fig1: DeltaOmega}
\end{figure}

For comparison, the monochromatic Rabi frequency with $E_1$ and frequency $\omega_1$ is $\Omega_{1} = e E_1 |\mel{\mathbb{1}}{x}{\mathbb{2}}|/\hbar$, where the dipole matrix elements $\mel{\mathbb{1}}{x}{\mathbb{2}}$ reflect the spin transition inside the qubit subspace. 
Without shear and inhomogeneous strains, for $E_{\text{gate}} = 10$~MV/m and $E_1=E_{\text{AC}} = 10$~kV/m (parallel to $B_x = 1$~T), we obtain $f_{\text{Rabi},1} = \Omega_{1}/2\pi = 10$~MHz at {$\omega_1 = \omega_0 =2.4\times2\pi$~GHz}, consistent with full scale atomistic tight-binding prediction and experiment \cite{Hendrickx2024,Abhikbrata2025}. 
In this regime, the AC Stark shift inside the qubit subspace manifests as Rabi splitting rather than a static shift; the corresponding second order shift is $\delta\omega^{(2)} = 0.3\times2\pi$~MHz.
The single qubit average gate fidelity (relative to $\mathrm{X}_\pi$ gate) is $\mathcal{F} = 99.95\%$, which still falls short of fault-tolerant thresholds, and charge noise dephasing will further reduce this value \cite{Veldhorst2015,Yoneda2018,Huang2019,Jiaan2024,Jiaan2025}.
The evaluation of this average gate fidelity can be found in the Supplementary Material \cite{supp}.

In the simplest case, the two drivings are near degenerate, i.e., $\omega_1 \approx \omega_2 \approx \omega_0$, the second order shift cannot be eliminated, and the phase difference becomes important.
The Rabi frequency $\Omega = |\Omega_1 e^{i\phi_1}+\Omega_2 \mathrm{e}^{\mathrm{i}(\phi_2+(\omega_2-\omega_1)t)}|$ exhibits an envelope at the beat frequency $|\omega_2-\omega_1|$.
If the measurement averages over many beat periods (i.e., does not resolve the envelope), the effective experimental signal is $(\Omega_1^2+\Omega_2^2)^{1/2}$.
If the beat is not averaged out ($\omega_2-\omega_1) t\ll1$ during the measurement, it is $(\Omega_1^2+\Omega_2^2 + 2\Omega_1 \Omega_2 \cos(\phi_2-\phi_1))^{1/2}$.
In practice, it is feasible for gate deactivation by engineering the relative phase or amplitude (e.g., vector I/Q control) so that the resonant components destructively interfere in the near-degenerate case \cite{Tian2000,Marko2024}, but it is beyond our scope.
For the non-degenerate case $\omega_1 \neq \omega_2$, the two tone drive generally produces quasiperiodic beat-note transients; thus, even though cancellation of $\delta \omega^{(2)}$ remains possible, efficient EDSR is not guaranteed.
In a Floquet picture, sizeable Rabi oscillations require a near-resonant channel $n \omega_1+m \omega_2 \simeq \omega_0$ ($n, m \in \mathbb{Z}$), while higher order channels are suppressed by detuning.
As an off-resonant estimate, the lowest mixed $(1,1)$ channel scales as
\begin{align}
\Omega_{(1,1)} \simeq \frac{\Omega_1 \Omega_2}{2}\left(\frac{\omega_1}{\omega_0^2-\omega_1^2}+\frac{\omega_2}{\omega_0^2-\omega_2^2}\right) \,.
\end{align}
This Floquet-Magnus type estimate applies for $\left|\omega_0-\omega_{1,2}\right| \gg \Omega_{1,2}$; its divergence as $\omega_1 \rightarrow \omega_0$ indicates breakdown near resonance, where a degenerate Floquet or Rabi treatment is required \cite{Tuorila2010}.

In Fig.~\ref{Fig1: DeltaOmega}, we plot $\delta\omega^{(2)}$ as a function of the driving frequencies $\omega_1$ and $\omega_2$.
The qubit Larmor frequency in this figure is $\omega_0 = 2.4\times 2\pi$~GHz.
Although a broad region with vanishing $\delta\omega^{(2)}$ exists, only the strongly resonant driving region (the green band) maintains a fast EDSR Rabi rate, which sets the scope of our attention in the following discussion.

The most relevant case is the primary driving at $\omega_1 = \omega_0$ with $E_1 = 10$~kV/m, while $\omega_2$ acts as an auxiliary tone to cancel the second order shift, i.e., $\delta \omega^{(2)} = 0$.  
This can be summarized by defining:
\begin{equation}
R_0 \equiv \frac{C_{\text{AC,1}}(\omega_2) + C_{\text{BS}}(\omega_2) + C_{\text{AC,2}}(\omega_2)}{C_{\text{BS}}(\omega_1) + C_{\text{AC,2}}(\omega_1)} = -\frac{E_1^2}{E_2^2} \,.
\end{equation}
However, $R_0$ should not be too large or too small, otherwise the required $E_1$ or $E_2$ becomes impractical for typical EDSR hardware.

\begin{figure}[htbp!]
\centering
\includegraphics[width=\linewidth]{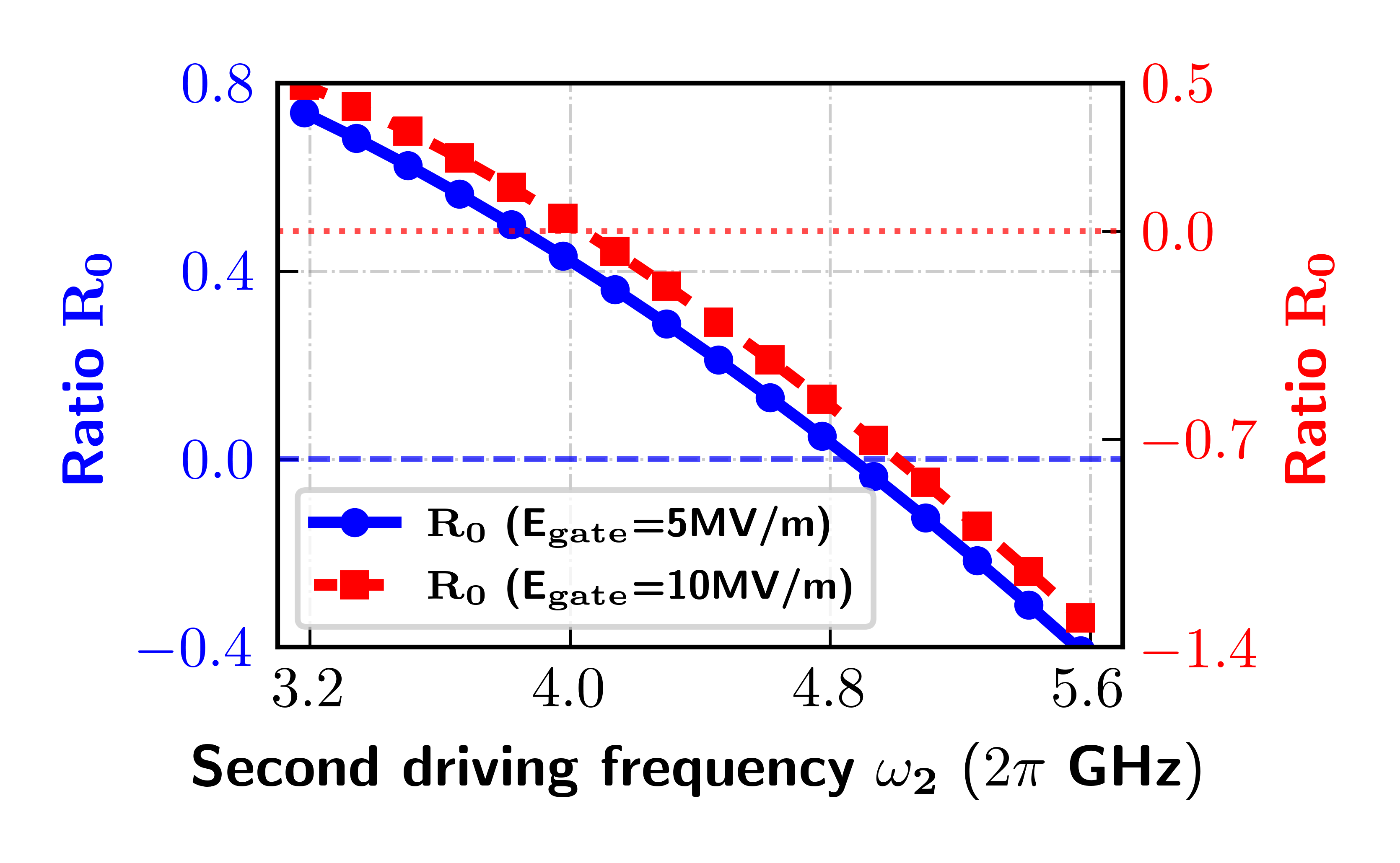}
\caption{$R_0$ versus the auxiliary frequency $\omega_2$ at $\omega_1=\omega_0$.
Two dashed lines indicate the frequency thresholds for $R_0<0$, which set the minimum auxiliary frequency.
Left y-axis: blue solid curve with circles, $E_{\text{gate}}=5$~MV/m, right y-axis: red dashed curve with squares, $E_{\text{gate}}=10$~MV/m.
The in-plane magnetic field is $B_x=1$~T.}
\label{Fig2: Ratio_Resonance}
\end{figure}

In Fig.~\ref{Fig2: Ratio_Resonance}, the ratio factor $R_0$ (for different $E_{\text{gate}}$, hence different $\omega_0$), changes sign as $\omega_2$ is swept.
A broad $\omega_2$ range can lead to {$\delta\omega^{(2)}=0$}, which then fixes the auxiliary amplitude $E_2 = E_1/\sqrt{R_0}$.
In this regime, the EDSR Rabi rate remains $\Omega_1$; the second driving contributes only sidebands that time-average to zero at the central resonance.
Choosing $\omega_2$ such that $R_0\leq-1$, allows $E_2<E_1$, reducing dissipation and crosstalk.  
However, we emphasize again that $\omega_2$ should avoid integer or half integer multiples of $\omega_0$ to prevent multiphoton processes, and satisfy $\Omega_{\alpha} \ll |\omega_2 - \omega_0|$ to avoid Autler-Townes splitting \cite{Wankai2025}.

Another advantage of bichromatic driving is the ability to compensate static resonance offsets defined as $\delta\omega_{\text{res}} = \delta\omega_c + \delta\omega^{(2)}$, where $\delta\omega_c$ is the quasi-static charge induced frequency shift, and $\delta\omega^{(2)}$ is the drive-induced second order shift.
We note $\delta\omega_{\mathrm{res}}$ characterizes the residual detuning under fixed-frequency control, rather than the quasi-static inhomogeneous dephasing rate.
Charge noise is often modelled as random telegraph noise (RTN) source from individual traps, whose ensemble produces a $1/f$ spectrum.
Here, we consider a single charge defect described by screened Thomas-Fermi potential $V_{\text{TF}}$ \cite{Davies1997,Culcer2010,Bermeister2014}, where $\hbar\delta\omega_c \equiv \mel**{\mathbb{2}}{V_{\text{TF}}}{\mathbb{2}}-\mel**{\mathbb{1}}{V_{\text{TF}}}{\mathbb{1}}$.
The residual detuning $\delta\omega_{\text{res}}$ directly produces detuning-induced gate errors, whose minimization can reduce recalibration overhead \cite{Takeda2018,Capannelli2025}.

Although we set the primary drive at $\omega_1=\omega_0$ to meet the one-photon EDSR resonance and maintain a large Rabi rate $\Omega_1$ \cite{supp}, Rabi oscillations can still decay due to slow charge induced shifts of the qubit splitting.
In the rotating frame this residual detuning tilts the rotation axis and leads to detuning-induced control errors.
The auxiliary tone at $\omega_2$ is chosen sufficiently detuned $(\Omega_\alpha \ll\left|\omega_2-\omega_0\right|)$, so it primarily dresses the qubit via virtual processes and generates an adjustable second order longitudinal shift $\delta \omega^{(2)}\left(\omega_2\right) \propto E_2^2$.
Since this shift depends on the same $E_{\text{gate}}$ working point, as $\omega_0, \omega_2$ can be tuned to minimize the noise sensitivity of the dressed splitting, thereby reducing the gate infidelities without sacrificing the Rabi frequency.

\begin{figure}[htbp!]
\centering
\includegraphics[width=\linewidth]{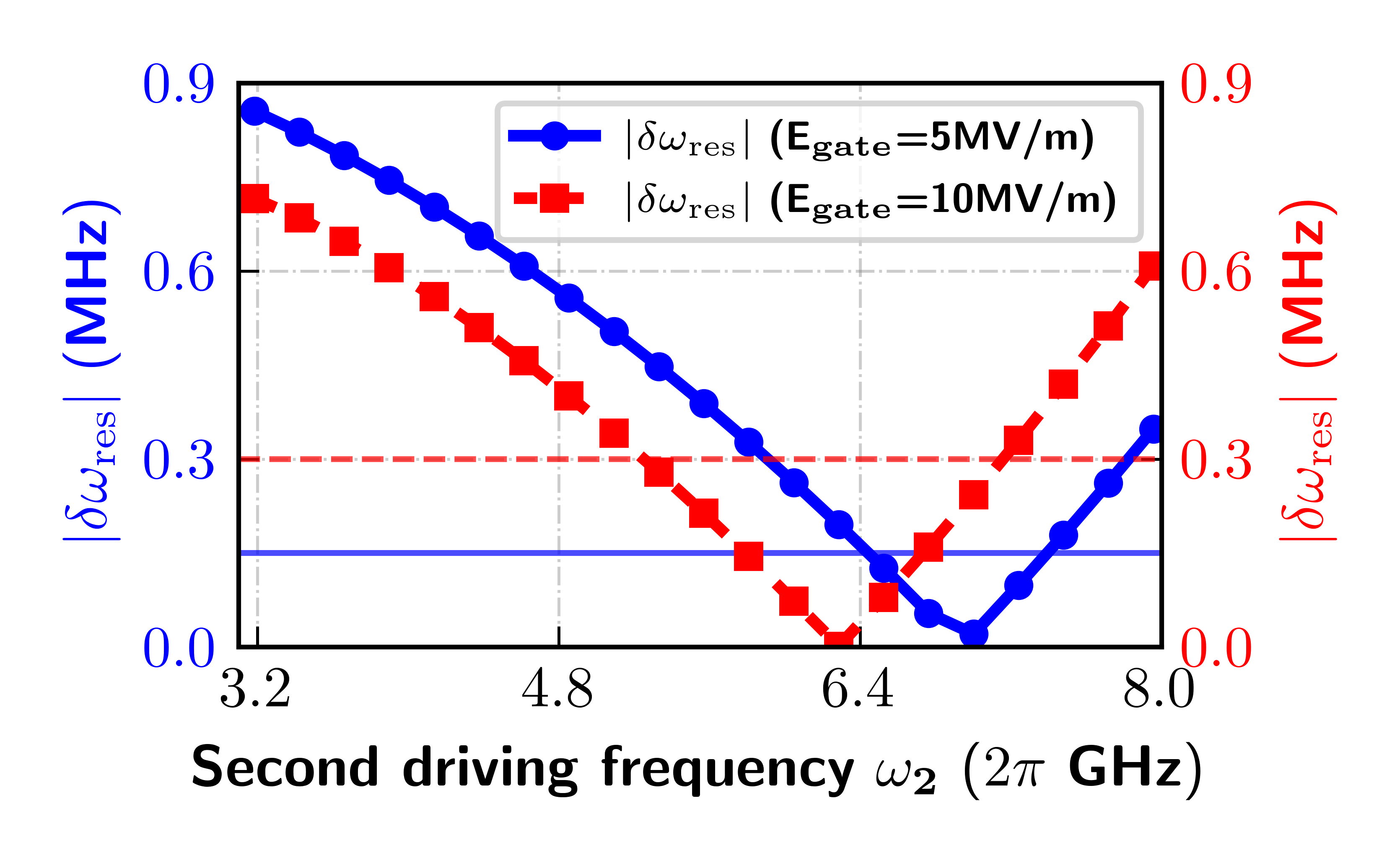}
\caption{Residual detuning magnitude $|\delta\omega_{\mathrm{res}}|$ versus $\omega_2$ under bichromatic driving at $\omega_1=\omega_0$ to maintain a fast Rabi rate.
Blue solid circles: $E_{\text{gate}}=5$~MV/m; red dashed squares: $E_{\text{gate}}=10$~MV/m.
The horizontal lines indicate the corresponding single-tone values of $|\delta\omega_{\mathrm{res}}|/2\pi$ at $\omega_1=\omega_0$ with $E_1$; the solid blue (dashed red) line corresponds to $E_{\text{gate}}=5$ ($10$~MV/m), respectively.
Parameters are taken from Ref.~\citenum{Abhik2023}.
}
\label{Fig3: T2vsOmega2}
\end{figure}

In Fig.~\ref{Fig3: T2vsOmega2}, we report the $|\delta\omega_{\text{res}}|$ under bichromatic driving as a function of $\omega_2$ for different $E_{\text{gate}}$.
We fix primary driving frequency $\omega_1 = \omega_0$ (resonance condition) to maintain a large EDSR Rabi rate.
To limit power dissipation, we sweep up to $\omega_2/2\pi \leq 8$ GHz ($\omega_2 \leq 50$~GHz).
The non-monotonic dependence of $|\delta\omega_{\text{res}}|$ on $\omega_2$ reflects the competition between the quasi-static charge induced offset and the auxiliary tone induced dispersive longitudinal shift $\delta\omega^{(2)}$.
For small $\omega_2$, $\delta\omega_c$ and $\delta\omega^{(2)}$ have the same sign so that $|\delta\omega_c + \delta\omega^{(2)}|$ increases, leading to a larger residual detuning.
As $\omega_2$ is increased further, $\delta\omega^{(2)}$ changes sign and a compensation point $|\delta\omega_c + \delta\omega^{(2)}| \approx 0$ is reached, where $|\delta\omega_{\text{res}}|$ is minimized.
Beyond this point, $\delta\omega^{(2)}$ no longer compensates $\delta\omega_c$, so that $|\delta\omega_c + \delta\omega^{(2)}|$ increases again, producing the rise on the other side of the minimum.

In conclusion, we have shown that bichromatic driving suppresses the second-order drive-induced frequency shift and reduces the residual detuning under fixed-frequency control in Ge hole spin qubits.
The scheme requires only minimal changes to standard EDSR techniques and is readily applicable to Si platforms, where the smaller Zeeman splitting permits lower auxiliary frequencies \cite{Peihao2021,Liles2021,Guangchong2025}.
It should also generalize to devices incorporating strain profiles, where enhanced SOC and larger dipole matrix elements are expected.
While our analysis focuses on the quasi-static limit, extending the Floquet framework can capture richer dynamics such as multiphoton processes, universal dynamical sweet spots, and Raman pathways \cite{Didier2019,Ziwen2021,Valery2022}.

\textit{Supplementary Material}.
In the Supplementary Material, we provide details on the Hamiltonian construction, gate fidelity estimation, and several supporting figures discussing the second order frequency shift in the monochromatic driving case.

\textit{Acknowledgements}.
We are very grateful to Zhihai Liu and the two anonymous reviewers for their valuable suggestions and constructive comments on this manuscript.

\textit{Data Availability}.
The data that support the findings of this study are available from the corresponding author upon reasonable request.

\end{document}